\documentclass[preprint,nofootinbib]{revtex4}%
\usepackage{amssymb}
\usepackage{cancel} 
\usepackage{amsfonts}
\usepackage{amsmath}
\usepackage{amsmath}
\usepackage{graphicx}
\usepackage{hyperref}
\global\long\def\dd{\mathrm{d}}
\usepackage[usenames]{color}%
\providecommand{\U}[1]{\protect\rule{.1in}{.1in}}
\hypersetup{colorlinks,linkcolor={blue},citecolor={blue},urlcolor={black}} 
\providecommand{\U}[1]{\protect\rule{.1in}{.1in}}
\definecolor{blue}{rgb}{0,0,1}

\newcommand{\ri}{ \mathrm{i}}
\definecolor{red}{rgb}{1,0,0}

\begin{document}
\title{Logarithmic corrections to the entropy of near-extremal black holes in Einstein-Gauss-Bonnet}
\author{Alejandro Alvarado$^1$, Andrés Anabalón$^{2,3}$, Mariano Chernicoff$^1$, Julio Oliva$^2$, Marcelo Oyarzo$^{4}$, Gabriel Ortega$^2$ and Jorge Urbina$^2$}
\affiliation{$^{1}$Departamento de Física, Facultad de Ciencias, Universidad
Nacional Autónoma de México, A.P. 70-542, CDMX 04510, México}
\affiliation{$^{2}$Departamento de F\'{\i}sica, Universidad de Concepci\'{o}n, Casilla,
160-C, Concepci\'{o}n, Chile}
\affiliation{$^{3}$Instituto de Física Teórica, UNESP-Universidade Estadual Paulista, R. Dr. Bento T. Ferraz 271, Bl. II, Sao Paulo 01140-070, SP, Brazil.}
\affiliation{$^{4}$Instituto Galego de Física de Altas Enerxías (IGFAE), Universidade de Santiago de Compostela,
E-15782 Santiago de Compostela, Spain.}
\begin{abstract}
We compute the one-loop contribution to the semiclassical partition function of near-extremal, asymptotically AdS black holes in five-dimensional Einstein-Gauss-Bonnet gravity. In the absence of an exact analytic rotating solution at finite Gauss-Bonnet coupling $\alpha$, we restrict to static, charged configurations and evaluate the contribution to $Z_{\text{1-loop}}$ arising from tensor, vector, and $U(1)$ gauge fluctuations. The analysis is based on the spectrum of a generalized Lichnerowicz operator governing linearized perturbations on the near-horizon geometry of the extremal solution, including its deformation by the coupling $\alpha$. In the canonical ensemble, the low-temperature behavior of the one-loop partition function leads to logarithmic corrections to the entropy of the form $\log(T/T_0)$, where the scale $T_0$ depends on both the fluctuation sector and the Gauss-Bonnet coupling. These corrections are controlled by the structure of zero modes of the deformed operator and their splitting at small but finite temperature. Our explicit computation yields a universal low-temperature scaling $Z_{\text{1-loop}}\sim 5 \log T$, where the coefficient arises from the combined contributions of tensor, vector, and $U(1)$ gauge modes, reflecting the corresponding counting of zero modes in each sector.
\end{abstract}

\maketitle

\section{Introduction}
Higher-curvature corrections to General Relativity are ubiquitous in schemes where the Einstein-Hilbert action emerges as the first term of an effective field theory in a low-energy expansion (see e.g. \cite{Goroff:1985sz,Zwiebach:1985uq,Gross:1986iv,Metsaev:1987bc,Metsaev:1987zx,Bergshoeff:1989de,Donoghue:1994dn}). In perturbative frameworks, the freedom to perform field redefinitions can be exploited to recast the theory into equivalent forms, thereby making the well-behaved dynamical properties more transparent. In particular, in the context of String Theory, when a quadratic term in the curvature is present, the Gauss-Bonnet combination naturally emerges. On its own merit, this quadratic combination originally appeared as the first deformation of General Relativity leading to second-order field equations on a generic background, namely as the second term in the Lovelock series \cite{Lovelock:1971yv}. Due to its simplicity, this framework provides a perfect model to study the physical effects induced by the higher-curvature corrections to General Relativity, including for example finite 't Hooft coupling effects in holography \cite{Buchel:2009sk,Kats:2006xp,Guica:2005ig,Brigante:2007nu,Brigante:2008gz,Camanho:2009hu,Camanho:2009vw,Hofman:2009ug,Camanho:2014apa}, exact black hole solutions, perturbation theory and hyperbolicity \cite{Boulware:1985wk,Wheeler:1985nh,Wheeler:1985qd,Corral:2022udb,Corral:2024xfv,Dotti:2004sh,Dotti:2005sq,Gleiser:2005ra,Choquetbruhat:1988jdt,Andrade:2016yzc,Kovacs:2020ywu}, and beyond (for a list of references see the early reviews \cite{Garraffo:2008hu,Charmousis:2008kc} and references therein and thereof).

We extend this program by explicitly computing the one-loop partition function contributions from gravitational tensor, vector, and $U(1)$ gauge modes. This yields a $\log T$ correction to the entropy of asymptotically Anti-de Sitter, near-extremal charged black holes in five-dimensional Einstein–Gauss–Bonnet gravity. The computation is motivated by the fact that the relevant operator, whose spectrum controls the one-loop corrected partition function is a deformation of the Lichnerowicz operator, which in this case depends on the Gauss-Bonnet coupling constant $\alpha$. As in General Relativity \cite{Iliesiu:2022onk,Karan:2022dfy,Banerjee:2023quv, Kapec:2023ruw,Rakic:2023vhv,Maulik:2024dwq,Maulik:2025phe,Blacker:2025zca,Acito:2025hka,PandoZayas:2026vbg}, the one-loop partition function of massless modes on the extremal black hole background diverges since the relevant operator $\mathbb{O}$ possesses zero-modes. Such a divergence can be regularized by heating up the black hole; the corresponding perturbed eigenvalues of $\mathbb{O}$ on the near-extremal black hole, after regularization, lead to a $\log{T}$ contribution to the one-loop partition function and, consequently, to a $\log{T}$ correction to the semiclassical gravitational entropy. The latter does not fulfill the area law, due to the presence of the dimensionful Gauss-Bonnet coupling $\alpha$ \cite{Jacobson:1993xs}\footnote{This entropy also naturally emerges as one of the charges associated to near-horizon symmetries in the presence of the Gauss-Bonnet term \cite{Chernicoff:2025dwl}.}. Following the fruitful approach taken in GR, instead of working with the whole spacetime, we take advantage of the decoupling limit induced by the small temperature regime, and work on the near-horizon geometry, which in the extremal case is locally given by the direct product of AdS$_2$ and $S^3$, with radii depending on the Gauss-Bonnet coupling and a $U(1)$ gauge field with field strength proportional to the volume-form of the AdS$_2$ factor. Finite temperature effects induce a linear in $T$ correction to both, the geometry and the gauge field. As shown below, the one-loop partition function in EGB, under these conditions reads
\begin{equation}\label{resultado}
\log{Z^{(0)}_\text{1-loop}}=\frac{3}{2}\log{\frac{T}{T_{\rm tensor}}}+\frac{6}{2}\log{\frac{T}{T_{\rm vector}}}+\frac{1}{2}\log{\frac{T}{T_{U(1)}}}+\ldots\ ,
\end{equation}
which at low temperature has the leading contribution $5\cdot\log{T}$ with a $3\cdot 1/2$ arising from the gravitational tensor modes, a $6\cdot 1/2$ contribution coming from the vector modes which use as building blocks the six Killing vectors of $S^3$, and a $1/2$ contribution of the $U(1)$ gauge modes, namely $5=3\cdot\frac{1}{2}+6\cdot\frac{1}{2}+1\cdot\frac{1}{2}$. Ellipsis in \eqref{resultado} stands for numerical values.  The quantities $T_{\rm tensor}, T_{\rm vector}, T_{U(1)}$ are mode-dependent constants which depend in turn on the parameters of the theory and the horizon radius of the extremal black hole.

The paper is organized as follows: In Section II, we present the theory under consideration and review the charged black holes that admit an extremal limit. We compute their thermodynamical variables and provide a realization, in the context of Einstein-Gauss-Bonnet coupled to Maxwell, of the Hawking radiation problem at tree level below the energy $E_{\rm gap}(\alpha)$. In Section III, in the canonical ensemble, we perform the near-horizon and near-extremal limit of the geometry presented in the previous section.  In Section IV, we define the generalized Lichnerowicz operator by computing functional derivatives of the action with Dirichlet boundary conditions on the metric and Neumann boundary conditions on the gauge fields. Following this result, we compute the zero-modes on the $T=0$ geometry, which consist of tensor and vector modes, classified by their transformation properties under diffeomorphisms on the $\rm AdS_2$ space, and $U(1)$ gauge modes due to the presence of the Maxwell field. We then lift these zero-modes by considering a non-vanishing but infinitesimal temperature in the background, computing the corresponding corrections that lead to the $\log T$ contribution to the entropy. In Section V, we present our concluding remarks.

\vspace{0.9em}
\noindent \textbf{Conventions } For Lorentzian spaces we use the mostly-plus convention. The Hodge star acting on a $p$-form, $F_p = \frac{1}{p!}F_{\mu_1 \dots \mu_p}\dd x^{\mu_1} \wedge \dots \wedge \dd x^{\mu_p}$ on a $D$-dimensional space with metric $g_{\mu \nu}$ and determinant $g$ is defined as $\star F_p = \frac{\sqrt{|g|}}{p!(D-p)!}F^{\mu_1 \dots \mu_p} \epsilon_{\mu_1 \dots \mu_p \nu_1 \dots \nu_{D-p}} \dd x^{\nu_1} \wedge \dots \wedge \dd x^{\nu_{D-p}}$ with $\epsilon_{x^1 \dots x^{D}} = 1$.

\vspace{0.5em}
\noindent\textbf{Public code } The results presented in Section IV for tensor, vector, and U(1) gauge modes can be reproduced using a Maple script provided at 

\hfill
\url{https://github.com/moyarzoca/egb-log-corrections-brackets}.

\section{The charged black hole in Einstein-Gauss-Bonnet}
The five-dimensional Einstein-Gauss-Bonnet theory coupled to a Maxwell field can be defined via the bulk action
\begin{align}
I_\text{bulk}[g,A] & =\frac{1}{\kappa^{2}}\int\dd^{5}x\sqrt{-g}\left(R-2\Lambda+\alpha\mathcal{G}-F_{\mu\nu}F^{\mu\nu}\right)\,,\\ \label{thetheory}
\mathcal{G} & =R^{\mu\nu\rho\sigma}R_{\mu\nu\rho\sigma}-4R^{\mu\nu}R_{\mu\nu}+R^{2}\,, 
\end{align}
with $\kappa^2 = 16 \pi G$. This action leads to the Maxwell equations, $\dd\star F=0$ and to
\begin{align}
G_{\mu\nu}+\Lambda g_{\mu\nu}+\alpha H_{\mu\nu} & =2\left(F_{\mu\lambda}F_{\nu}{}^{\lambda}-\frac{1}{4}g_{\mu\nu}F_{\lambda\delta}F^{\lambda\delta}\right)\,,
\end{align}
where the Gauss-Bonnet tensor $H_{\mu\nu}$ is defined via the functional derivative
\begin{align}
H_{\mu\nu} & =\frac{1}{ \sqrt{-g}}\frac{\delta}{\delta g_{\mu\nu}}\int\dd^{5}x\sqrt{-g}\mathcal{G}\,,\\
 & =2RR_{\mu\nu}-4R_{\mu\rho}R^{\rho}{}_{\nu}-4R^{\delta}{}_{\rho}R^{\rho}{}_{\mu\delta\nu}+2R_{\mu\rho\sigma\delta}R_{\nu}{}^{\rho\sigma\delta}-\frac{1}{2}\mathcal{G}g_{\mu\nu}\,.
\end{align}
This theory admits a static electrically charged solution of the form \cite{Wiltshire:1985us}
\begin{equation}
\dd s^{2} =-f(r)\dd t^{2}+\frac{\dd r^{2}}{f(r)}+r^{2}\dd s^{2}(S^{3})\,,
\end{equation}
with metric function\footnote{The metric function (\ref{f_of_r}) is also a solution of (\ref{thetheory}) with a plus sign in front of the square root. However, in the limit $\alpha\rightarrow0$, this is not a solution of General Relativity. We will focus on the Einstein branch which is perturbativelly connected with the five-dimensional, charged Schwarzschild-Tangherlini solution.} and one-form gauge field
\begin{align}
f(r) & =1+\frac{r^{2}}{4\alpha}\left(1-\sqrt{1+\frac{4\alpha\Lambda}{3}+\frac{m\alpha}{r^{4}}-\frac{16q^{2}\alpha}{3r^{6}}}\right) \label{f_of_r},\\
A & =\frac{1}{\sqrt{2}}\left(\frac{q}{r^{2}}-\frac{q}{r_+^2}\right)\dd t\,.
\end{align}
Here, $m$ is proportional to the black hole mass, and $q$ is proportional to the electric charge, while $r_+>0$ stands for the event horizon located at the largest zero of the function $f(r)$. In the limit $\alpha\rightarrow 0$, the charged black hole reduces to the charged Schwarzschild-Tangherlini solution of the Einstein-Maxwell theory in five dimensions. Notice also that, as $r\rightarrow\infty$ the spacetime is asymptotically flat when $\Lambda$ vanishes, or asymptotically (A)dS with an effective cosmological constant, since
\begin{equation}
f(r)=\frac{r^2}{\ell_\text{eff}^2}+1-\frac{3m}{8\sqrt{9+12\alpha\Lambda}r^2}+\frac{2q^2}{\sqrt{9+12\alpha\Lambda}r^4}+\mathcal{O}(r^{-6})\ ,
\end{equation}
where the dressed AdS radius is determined via
\begin{equation}
   \ell_\text{eff}^{2} = -\frac{3}{\Lambda}\left( 1 + \sqrt{1 + \frac43 \alpha \Lambda} \right)\, .
\end{equation}
Since we are eventually interested in the emergence of the Gauss-Bonnet term from String Theory $\alpha'$ corrections, we disregard the case $4\Lambda\alpha=-3$, where the action can be written as a Chern-Simons theory for the AdS group \cite{Chamseddine:1989nu}.

The black objects for $\alpha\geq0$ possess event and Cauchy horizons that surround a singularity at $r=0$. One could potentially have a branch singularity at the value of $r$ at which the radicand in \eqref{f_of_r} vanishes. For a thorough analysis of the possible causal structures, including the topological extensions of charged black holes with (non-)vanishing cosmological constant, see \cite{Torii:2005nh}. \\

A well-defined action principle requires supplementing \eqref{thetheory} with appropriate boundary terms. Specifically, to ensure a well-defined variational principle, we include the generalized Gibbons–Hawking–York (GHY) term \cite{Gibbons:1976ue,Myers:1987yn}. Furthermore, to guarantee a finite action in asymptotically AdS$_5$ spacetimes, local counterterms must be added \cite{Emparan:1999pm,Brihaye:2008xu,Anastasiou:2025usa}. These are respectively given by:
\begin{align}
I_{\mathrm{GHY}} &=\frac{1}{\kappa^{2}}\int_{\partial M} \dd^{4}x\sqrt{-h}\left[2K+\alpha\delta_{ijk}^{abc}K^{i}{}_{a}\left(\frac{1}{2}\mathcal{R}^{jk}{}_{bc}-\frac{1}{3}K^{j}{}_{b}K^{k}{}_{c}\right)\right]\,,\\
I_{\mathrm{ct}} &= \frac{2}{\kappa^2} \int \dd^4 x \sqrt{-h} \left[ -\frac{1}{\ell_{\rm eff}}\left(2+ \sqrt{1-\frac{8 \alpha}{L^2}} \right) - \frac{\ell_{\rm eff}}{4}\left(2- \sqrt{1-\frac{8 \alpha}{L^2}} \right) \mathcal{R} \right]\,,\label{Ict}
\end{align}
where $\Lambda=-6/L^2<0$. Additionally, to implement the canonical ensemble (i.e., for fixed temperature and charge in AdS), the action is further supplemented by the Hawking–Ross term \cite{Hawking:1995ap}:
\begin{equation}
    I_{\mathrm{HR}}= \frac{4}{\kappa^2}\int_{\partial \mathcal{M}} \dd^4 x \sqrt{-h}\, n_aF^{ab}A_b = \frac{4}{\kappa^2} \int_{\partial \mathcal{M}} A \wedge \star F\,.
\end{equation}\\

For future convenience, it is useful to express some quantities in terms of the location of the inner and the outer horizons. To this end, we first rewrite \eqref{f_of_r} as a function of $r_+$ and $r_-$ as
\begin{equation}
f(r) = 1+\frac{r^2}{4\alpha} \left[ 1 - \sqrt{ \left(1 + \frac{4\alpha}{r^2}\right)^2 + \frac{4\alpha\Lambda}{3r^6}(r^2-r_+^2)(r^2-r_-^2)\left(r^2 -\frac{6}{\Lambda} + r_+^2 + r_-^2\right) }\right]\ \label{frmasrmenos} ,
\end{equation}
where we have used the expressions
\begin{equation}
    m = 16\alpha +8(r_+^2+r_-^2)-\frac{4\Lambda}{3}(r_+^4 +r_+^2r_-^2 +r_-^4)\ ,\label{mu_parameter}
\end{equation}
and
\begin{equation}
     q^2=\frac{3}{2} r_+^2 r_-^2 -\frac{\Lambda}{4} r_+^2 r_-^2 (r_+^2 + r_-^2)\ . \label{q_as_function_of_rm_rp}
\end{equation}
Using \eqref{frmasrmenos}, the temperature of the black hole reads
\begin{equation}
    T= \frac{f'(r_+)}{4\pi}=\frac{(r_+^2 -r_-^2)(6-2\Lambda r_+^2  -\Lambda r_-^2 )}{12 \pi r_+(r_+^2 + 4\alpha)}.
\end{equation}
It is possible to compute the entropy of the black hole using the Wald's formula \cite{Wald:1993nt, Iyer:1994ys} of a horizon $\mathcal{H}$. Let $\xi$ be a horizon generator, the Wald's entropy for the theory \eqref{thetheory} is given by
\begin{align}
S & =\frac{1}{T}\int_{\mathcal{H}}\star\mathbf{q}[\xi]\,,\\
\mathbf{q} & =-\frac{1}{\kappa^{2}}\left[\nabla_{\rho}\xi_{\sigma}\left(1+2\alpha R\right)+2\alpha\nabla^{\mu}\xi^{\nu}R_{\rho\sigma\mu\nu}-8\alpha\nabla^{\mu}\xi_{\sigma}R_{\rho\mu}\right]\dd x^{\rho}\wedge\dd x^{\sigma}\,.
\end{align}
For the static black hole, one can just consider $\xi = \partial_t$, then the entropy for the outer horizon located at $r_+$ is
\begin{equation}
    S = \frac{\pi^2 r_{+}^3}{2 G}\left(1 + \frac{12 \alpha}{r_+^2} \right) \, .
\end{equation}
Notice that this entropy does not fulfill the area law, since both $r_+^3$ and $\alpha r_+$ have the same dimension and contribute to it. With this normalization, it is possible to determine the charge $Q$, the chemical potential $\Phi$, and the mass $M$ of the configuration such that the first law of thermodynamics, $ \dd M = T \dd S + \Phi \dd Q$, is satisfied. These quantities are given by
\begin{align}
       Q &= \frac{4}{\kappa^2} \int \star F = - \frac{\pi q}{\sqrt{2} G} \, , \\
    \Phi &= - \frac{q}{\sqrt{2} r_+^2} \, ,  \\
    M &= \frac{3 \pi m}{64 G}\,. \label{mass}
\end{align}
Here after we will refer to the mass as $m$ or $M$ and to the charge as $q$ or $Q$, indistinctively, in order to have simpler expression.

For fixed values of the coupling $\alpha$ and the charge $Q$, there exists a minimum mass $M=M_{\text{ext}}(Q)$ for which the inner and the outer horizons coincide $r_0:=r_+=r_-$; in this case, the black hole becomes extremal. Due to the expression \eqref{mass}, this mass can be read from \eqref{mu_parameter} at extremality as
\begin{align}
m_{\rm ext} & = 16 r_0^2(q) +16 \alpha -4 r_0^4(q) \Lambda \,,
\end{align}
with
\begin{align}
    r_0^2(q) &= \frac{1}{\Lambda} \left[ 1 + 2 \cos\left( \frac{1}{3} \arccos(1 - q^2 \Lambda^2) + \frac{2n\pi}{3} \right) \right] \, ,\qquad \Lambda \neq 0 \, , \\ r_0^2(q) &= \sqrt{\frac23} |q| \,, \qquad \Lambda = 0\,,
\end{align}
where $n=0,1,2,\dots$, and $r_0$ is defined as the largest of these values. We work in the canonical ensemble, therefore $Q$ is arbitrary but fixed. A near-extremal solution is then characterized by a small temperature expansion. The energy can be formally written as $M=M(T,Q)$, and expanding around $T=0$, one obtains
\begin{align}
m & =m_{\mathrm{ext}}-\frac{16\pi^{2}(r_{0}^{2}+4\alpha)^{2}}{r_{0}^{2}\Lambda-2}T^{2}+\mathcal{O}(T^{3})\,,
\end{align}
namely
\begin{align}
M&=M_{\text{ext}}(Q) + \frac{T^2}{M_{\rm gap}}+\mathcal{O}(T^3)\ , 
\end{align}
with
\begin{equation}
M_{\rm gap}:= \frac{4 G (2- \Lambda r_0^2)}{3 \pi^3 (4 \alpha + r_0^2)^2} \, .
\end{equation}
As we are considering $\Lambda\leq 0$,  $M_{\rm gap}>0$. Consequently, as in General Relativity, deviations from extremality at finite temperature generate an $\mathcal{O}\left(T^2\right)$ correction to the energy above the extremal bound at fixed charge. Since the typical energy of a Hawking quantum is of order $T$, while the excess energy above extremality at fixed charge scales as $T^2$, there is insufficient energy to emit even a single quantum at sufficiently low temperatures. 
As in General Relativity, these results suggest that the density of states of the black hole is sharply peaked at the extremal groundstate, since the entropy of the extremal black hole can be arbitrarity large, and then such large microstate degeneracy is gapped with respect to the first radiating thermal hole. Such a density of states typically reflects a high degree of symmetry in the extremal configuration, as in some supergravity models. Yet, as in the Einstein–Maxwell vacuum, no corresponding symmetry is known that would explain a large, gapped ground-state degeneracy in Einstein-Gauss-Bonnet-Maxwell theory. This is a well-known puzzle in General Relativity \cite{Preskill:1991tb,Maldacena:1998uz,Page:2000dk}, and here we have argued that it is also present in the charged Einstein-Gauss-Bonnet black hole. In our case, $E_\text{gap}$ depends on the Gauss-Bonnet coupling $\alpha$ as
\begin{equation}
E_\text{gap}= \frac{4 G (2- \Lambda r_0^2)}{3 \pi^3 (4 \alpha + r_0^2)^2} \, ,
\end{equation}
which for perturbative $\alpha$ departs from its GR counterpart by a term that is linear in $\alpha$ itself. From the First Law in the Canonical Ensemble, considering two, infinitesimally close equilibrium configurations with the same charge, one has that $\delta M=T\delta S$, and it can be directly seen that in order to have a mass with a linear term in $T$ at low temperature, which may allow to emit Hawking quanta for arbitrarily low $T$, it must be that the entropy possesses a $\log T$ term. In General Relativity, it has been shown that such $\log T$ terms emerge when computing the one-loop, quantum correction to the semiclassical Bekenstein-Hawking entropy \cite{Iliesiu:2022onk,Karan:2022dfy,Banerjee:2023quv, Kapec:2023ruw,Rakic:2023vhv,Maulik:2024dwq,Maulik:2025phe,Blacker:2025zca,Acito:2025hka,PandoZayas:2026vbg}, and from the Euclidean functional integral at one-loop, one can compute the prefactor of the logarithmic term. This is what we do in what follows for the Einstein-Gauss-Bonnet theory, after characterizing the near-horizon, near-extremal configuration.

\section{Near-Horizon, Near-Extremal Geometry}
The Euclidean near-horizon geometry of the near-extremal solution is obtained by expanding the spacetime around the $T=0$ limit. To this end, we introduce the following coordinate transformation:
\begin{equation}
    r=r_+(T)+4\pi\ell_{\rm AdS}^2T\sinh^2{\left(\frac{\eta}{2}\right)}\quad,\quad t=-\frac{i\tau}{2\pi T}\ ,
\end{equation}
with
\begin{equation}\label{lAdS2}
    \ell_{\rm AdS}^2=\frac{r_0^2+4\alpha}{2(2-\Lambda r_0^2)}.
\end{equation}
In the strict $T\rightarrow 0$ limit, this transformation decouples an infinite AdS$_2$ throat with curvature radius $\ell_{\rm AdS}$ from the rest of the spacetime. At first order in the low-temperature expansion, the geometry describes the near-horizon region of a near-extremal black hole, such that:
\begin{equation}\label{NHNEXT}
g_{\mu\nu}=\bar{g}_{\mu\nu}+\delta g_{\mu\nu}T+\mathcal{O}(T^2)\ ,
\end{equation}
with the leading AdS$_2\times S^3$ spacetime given by
\begin{equation}\label{NHEXT}
    \bar{g}_{\mu\nu}\dd x^\mu\dd x^\nu = \dd s^2({\rm AdS_2}) + r_0^2\dd s^2({\rm S^3}) \, .
\end{equation}
Using the coordinates $(\tau, \eta)$ for the ${\rm AdS_2}$ space and $(\theta,\phi,\psi)$ for the $S^3$, the line elements in \eqref{NHEXT} are respectively
\begin{align}
    \dd s^2({\rm AdS_2}) &= \ell_{\rm AdS}^2(\sinh^2{\eta}\dd\tau^2+\dd\eta^2)\,, \label{metric-ads}\\
    \dd s^2({\rm S^3}) &= \frac14 (\dd \theta^2 + \sin^2 \theta \dd \phi^2) + \frac14 (\dd \psi + \cos \theta \dd \phi)^2\,. \label{metric-sphere}
\end{align}
The deformation at leading order in $T$ is given by
\begin{align}
    \delta g_{\mu\nu} dx^\mu dx^\nu &=2\pi \ell_{\rm AdS}^2\frac{\left[\ell_{\rm AdS}^2(6-r_0^2\Lambda)+3r_0^2\right]}{3r_0 (2-r_0^2\Lambda)}(2+\cosh\eta)\tanh^2\left(\frac{\eta}{2}\right)(-\sinh^2 \eta\, d\tau^2 +d\eta^2) \nonumber  \\ 
    &\qquad + 4\pi \ell_{\rm AdS}^2\, r_0  \cosh \eta \, d\Omega_3^2.
\end{align}
Note that this geometry depends on the Gauss-Bonnnet coupling $\alpha$ exclusively through the AdS$_2$ curvature radius $\ell_{\rm AdS}$ given in \eqref{lAdS2}, as the remaining coefficients remain insensitive to variations in $\alpha$ for fixed charge $q$. The expansion also acts on the gauge field, which takes the form $A^{\prime}=\overline{A}+\delta A \,T$, where
\begin{align}
\overline{A} & =\frac{\ell_{\mathrm{AdS}}^{2}}{r_{0}}\sqrt{3-r_{0}^{2}\Lambda}(\cosh\eta-1)\ri \dd \tau\,,\\
\delta A & =-\frac{3\pi\ell_{\mathrm{AdS}}^{4}}{r_{0}^{2}}\sqrt{3-\Lambda r_{0}^{2}}\sinh^{2}\eta\,\ri\dd \tau\,.
\end{align}
It can be verified that the field strength $\overline{F}=d\overline{A}$, associated with $\overline{A}$ is proportional to the volume form of the AdS$_{2}$ factor in \eqref{NHEXT}, while the background gauge field vanishes at the origin. 
Regularity at the origin, located at $\eta=0$, of the near-horizon extremal geometry \eqref{NHEXT}, requires $\tau \sim\tau+2\pi$, which in turn facilitates the mode decomposition of fields on this background.

The near-horizon geometry \eqref{NHNEXT} constitutes a perturbative solution of the Einstein-Gauss-Bonnet field equations coupled to a Maxwell field, valid up to $\mathcal{O}(T^2)$. In performing this expansion, we have imposed that along the line parameterized by $T$, the charge parameter $q$ in \eqref{q_as_function_of_rm_rp} remains fixed. This condition uniquely determines the locations of the event horizon $r_+$ and the Cauchy horizon $r_-$ as functions of $T$, which are given by
\begin{align}
    r_{+}(T) &= r_0+2\pi\ell_{\rm AdS}^2T+\pi^2\ell_{\rm AdS}^2\left(\frac{4 (6- r_0^2\Lambda )\alpha + 14 r_0^2-5\Lambda r_0^4}{r_0(2-\Lambda r_0^2)^2}\right)T^2 \nonumber \\
    & \quad +\frac{8 \ell_{\rm AdS}^2 \pi^3}{3r_0^2(2 - \Lambda r_0^2)^4} 
    [ (96 + 32 r_0^2 \Lambda - 16 r_0^4 \Lambda^2)\alpha^2 + (168 r_0^2 -68 r_0^4 \Lambda +4 r_0^6 \Lambda^2)\alpha \nonumber  \\
    & \quad + 48 r0^4 -31 r_0^6 \Lambda +5 r_0^8\Lambda^2
    ]\,T^3 +\mathcal{O}(T^4)\, ,
\end{align}

and
\begin{align}
    r_{-}(T) &= r_0-2\pi\ell_{\rm AdS}^2T-\pi^2\ell_{\rm AdS}^2\left(\frac{4(6- r_0^2\Lambda )\alpha +30r_0^2-13 r_0^4 \Lambda}{3 r_0(2-\Lambda r_0^2)^2}\right)T^2 \nonumber
    \\
    & \quad +\frac{4 \ell_{\rm AdS}^2 \pi^3}{9(2 - \Lambda r_0^2)^4} 
    [ (-384\Lambda +112 r_0^2 \Lambda^2)\alpha^2 +(-432 +120 r_0^2 \Lambda + 8 r_0^4 \Lambda^2) \alpha \nonumber \\ 
    & \quad -180 r_0^2 +126 r_0^4 \Lambda -23 r_0^6 \Lambda^2]\, T^3+\mathcal{O}(T^4)\, .
\end{align}
The above expressions will be essential for computing the  one-loop partition function around the near-horizon geometry of the near-extremal black holes.

\section{The generalized Lichnerowicz operator and its zero modes}
To compute the one-loop partition function on the near-horizon, near-extremal configuration, we expand the action \eqref{thetheory} up to second order in the perturbations $g_{\mu\nu}^{\prime}=g_{\mu\nu}+h_{\mu\nu}$ and $A_{\mu}^{\prime}=A_{\mu}+a_{\mu}$. The total action is decomposed as:
\begin{align}
I[g,A] & =I_{R}[g]+I_{\Lambda}[g]+I_{A}[g,A]+I_{\mathcal{G}}[g]\,,\label{splitting_terms_sum}
\end{align}
where:
\begin{align}
I_{R}[g] & =\frac{1}{\kappa^{2}}\int\dd^{5}x\sqrt{-g}R + \frac{1}{\kappa^2}\int_{\partial \mathcal{M}}  2 K \, \bar{\star} 1  \,, \label{IR_def}
\\
I_{\Lambda}[g]&=\frac{1}{\kappa^{2}}\int\dd^{5}x\sqrt{-g}(-2\Lambda)\,,\\
I_{A}[g,A] & =\frac{1}{\kappa^{2}} \int\dd^{5}x\sqrt{-g}(-F_{\mu\nu}F^{\mu\nu}) + \frac{4}{\kappa^{2}}\int_{\partial \mathcal{M}} A \wedge \star F\,, \label{IA_def}
\\
I_{\mathcal{G}}[g] &=\frac{1}{\kappa^{2}}\int\dd^{5}x\sqrt{-g} \ \alpha\mathcal{G} + \frac{1}{\kappa^2}\int_{\partial \mathcal{M}}\alpha \delta^{abc}_{def} K^{d}{}_a \left(\frac12 \mathcal{R}^{ef}{}_{bc} - \frac13 K^e{}_{b} K^f_{c} \right) \bar{\star} 1  \,, \label{IG_def}
\end{align}
Here, $\bar{\star}$ denotes the Hodge dual on $\partial \mathcal{M}$. Around a generic background the expansion reads
\begin{align}
&I[g+h,A+a]  =I[g,A]+\int\dd^{5}x\left(\frac{\delta I}{\delta g_{\mu\nu}(x)}h_{\mu\nu}(x)+\frac{\delta I}{\delta A_{\mu}(x)}\delta A_{\mu}(x)\right) \nonumber\\
 & \quad+\frac{1}{2}\int\dd^{5}x\int\dd^{5}y\left(h_{\mu\nu}(x)\frac{\delta^{2}I}{\delta g_{\mu\nu}(x)\delta g_{\rho\sigma}(y)}h_{\rho\sigma}(y)+a_{\mu}(x)\frac{\delta^{2}I}{\delta A_{\mu}(x)\delta g_{\rho\sigma}(y)}h_{\rho\sigma}(y)\right.\nonumber \\
 & \left.\quad+h_{\mu\nu}(x)\frac{\delta^{2}I}{\delta g_{\mu\nu}(x)\delta A_{\rho}(y)}a_{\rho}(y)+a_{\mu}(x)\frac{\delta^{2}I}{\delta A_{\mu}(x)\delta A_{\nu}(y)}a_{\nu}(y)\right)\,+\dots\ .\label{I_expanded}
\end{align}
The first-order variation of the action vanishes, as we are considering quantum fluctuation around a classical saddle point that satisfies Dirichlet boundary conditions for the metric and Neumann boundary conditions for the gauge field. The latter is required to implement the canonical ensemble. The first functional derivative of each term in \eqref{I_expanded} is given by: 
\begin{align}
\frac{\delta I_{R}}{\delta g_{\mu\nu}(x)} & =-\frac{\sqrt{-g}}{\kappa^{2}}G^{\mu\nu}\,,
\hspace{2cm}
\frac{\delta I_{\Lambda}}{\delta g_{\mu\nu}(x)}  =-\frac{\sqrt{-g}}{\kappa^{2}}\Lambda g^{\mu\nu}\,,
\\
\frac{\delta I_{A}}{\delta A_{\mu}(x)} & =\frac{4\sqrt{-g}}{\kappa^{2}}\nabla_{\rho}F^{\rho\mu}\,,
\hspace{1.7cm}
\frac{\delta I_{\mathcal{G}}}{\delta g_{\mu\nu}(x)}  =-\frac{\alpha\sqrt{-g}}{\kappa^{2}}H^{\mu\nu}\,,
\\
\frac{\delta I_{A}}{\delta g_{\mu\nu}(x)} & =\frac{\sqrt{-g}}{\kappa^{2}}\left(-\frac{1}{2}g^{\mu\nu}F_{\rho\sigma}F^{\rho\sigma}+2F^{\mu}{}_{\lambda}F^{\nu\lambda}\right) =: \frac{2\sqrt{-g}}{\kappa^{2}}T^{\mu\nu}\,.
\end{align}
Note that all the boundary contributions vanish by virtue of the imposed boundary conditions and the presence of the boundary terms in \eqref{IR_def}, \eqref{IA_def} and \eqref{IG_def}.
For simplicity, we shall treat the computation of each term in the sum (\ref{splitting_terms_sum}) separately. Let us introduce the notation
\begin{align}
I^{(f,g)} & := \frac12\int\dd^{5}x\int\dd^{5}y\,f_{M}(y)\frac{\delta^{2}I}{\delta F_{M}(y) \delta G_{N}(x)}g_{N}(x)\,,
\end{align}
for $F_M,G_M\in \{A_\mu, g_{\mu \nu}\}$ and $f_M,g_M \in \{ a_\mu, h_{\mu \nu} \}$.
The quadratic contributions to the action are, from the quadratic metric fluctuations associated with the cosmological constant:
\begin{align}
I_{\Lambda}^{(h,h)} & =\frac{1}{\kappa^{2}}\int\dd^{5}x\sqrt{-g}\,h_{\rho\sigma}\left(-\frac{\Lambda}{4}g^{\rho\sigma}g^{\mu\nu}+\frac{\Lambda}{2}g^{\mu\rho}g^{\sigma\nu}\right)h_{\mu\nu}\,,
\end{align}
while the quadratic metric fluctuations associated with the Maxwell term are:
\begin{align}
I_{A}^{(h,h)} & =\frac{1}{\kappa^{2}}\int\dd^{5}x\sqrt{-g}\,h_{\rho\sigma}\left[\frac{1}{8}F_{\lambda\delta}F^{\lambda\delta}\left(2g^{\rho\mu}g^{\sigma\nu}-g^{\rho\sigma}g^{\mu\nu}\right)-F^{\mu\rho}F^{\nu\sigma} + \right.\\
& \left. \vphantom{\frac{1}{1}}
\hspace{4cm}-2F^{\rho\lambda}F^{\mu}{}_{\lambda}g^{\sigma\nu}+g^{\mu\nu}F^{\rho}{}_{\lambda}F^{\sigma\lambda}\right]h_{\mu\nu}\,. \notag
\end{align}
As it is well-known, the quadratic metric fluctuations of the Einstein-Hilbert term is \cite{Christensen:1979iy}:
\begin{align}
I_{R}^{(h,h)} & =\frac{1}{\kappa^{2}}\int_{M}\dd^{5}x\sqrt{-g}\,h_{\alpha\beta}\left[\frac{1}{4}g^{\mu\alpha}g^{\nu\beta}\Box-\frac{1}{8}g^{\alpha\beta}g^{\mu\nu}\Box-\frac{1}{2}R^{\mu\alpha\beta\nu}+\frac{1}{2}g^{\beta\mu}R^{\alpha\nu}\right.\nonumber\\
 & \quad\left.-\frac{1}{2}g^{\mu\nu}R^{\alpha\beta}-\frac{1}{4}g^{\mu\alpha}g^{\nu\beta}R+\frac{1}{8}Rg^{\alpha\beta}g^{\mu\nu}\right]h_{\mu\nu}\nonumber\\
 & \quad+\frac{1}{32\pi G}\int\dd^{5}x\sqrt{-g}\,g^{\mu\nu}\left(\nabla_{\rho}h_{\mu}{}^{\rho}-\frac{1}{2}\nabla_{\mu}h\right)\left(\nabla_{\sigma}h_{\nu}{}^{\sigma}-\frac{1}{2}\nabla_{\nu}h\right)\label{Lich}\ .
\end{align}
The interaction terms coming from the functional derivative of the Maxwell equations preserving the boundary terms reads
\begin{align}
I_{A}^{(h,a)} & =\frac{1}{\kappa^{2}}\int\dd^{5}x\sqrt{-g}\,h_{\alpha\beta}\left(-g^{\alpha\beta}F^{\mu\nu}-4g^{\alpha\lambda}g^{\beta[\mu}F_{\lambda}{}^{\nu]}\right)\nabla_{\mu}a_{\nu} \\
 & \quad +\frac{1}{\kappa^{2}}\int\dd^{5}x\sqrt{-g}\,\nabla_{\rho}\left(h_{\alpha \beta}g^{\alpha\beta}F^{\rho\mu}a_{\mu}-4h_{\alpha\beta}g^{\alpha\lambda}g^{\beta[\rho}F_{\lambda}{}^{\mu]}\right) \, ,
\end{align}
while the interaction term from the functional derivative of the energy-momentum tensor is
\begin{align}
I_A^{(a,h)} & =\frac{1}{\kappa^{2}}\int\dd^{5}x\sqrt{-g}\,\nabla_{\lambda}a_{\sigma} (4g^{\mu[\sigma}F^{\lambda]\nu}-g^{\mu\nu}F^{\lambda\sigma})h_{\mu\nu}\,.
\end{align}
Note that $I_A^{(a,h)}=I_A^{(h,a)}$ for appropriate boundary conditions, leading to a factor of 2 in the interaction term. The Maxwell contribution for the gauge fluctuations leads to
\begin{align}
I_A^{(a,a)} & =\frac{1}{\kappa^{2}}\int\dd^{5}x\sqrt{-g}\,2a_{\mu}(g^{\mu\lambda}\Box-R^{\mu\lambda})a_{\lambda} +\frac{1}{\kappa^{2}}\int\dd^{5}x\sqrt{-g}\,2(\nabla^{\mu}a_{\mu})^{2}\,.\label{Maxwellaa}
\end{align}
Finally, one of the main results of this work is the quadratic fluctuation of the metric coming from the Gauss-Bonnet term, leading to
\begin{align}
I_{\mathcal{G}}^{(h,h)} & =\frac{1}{\kappa^{2}}\int\dd^{5}x\sqrt{-g}h_{\mu\nu}\left(-\frac{\alpha}{4}g^{\rho\sigma}H^{\mu\nu}+\alpha g^{\mu\rho}H^{\sigma\nu}-\frac{\alpha}{2}\mathsf{P}[H]^{\mu\nu,\rho\sigma}\right)h_{\rho\sigma}\,.
\end{align}
where the operator $\mathsf{P}[H]$ can be compactly presented as

\begin{align}
 & g_{\mu\lambda}g_{\nu\delta}\mathsf{P}^{\lambda\delta,\rho\sigma}h_{\rho\sigma}= \nonumber \\
 & -\frac{1}{2}h_{\mu\nu}\mathcal{G}-h^{\gamma\delta}\left[2R_{\gamma\delta}R_{\mu\nu}-4R_{\mu\gamma}R_{\nu\delta}-4R_{\gamma\eta\delta\lambda}R_{\mu}{}^{\eta}{}_{\nu}{}^{\lambda}+R_{\mu\gamma\eta\lambda}R_{\nu\delta}{}^{\eta\lambda}\right.+2R_{\mu}{}^{\eta}{}_{\gamma}{}^{\lambda}R_{\nu\lambda\delta\eta}\nonumber \\
 & \left.+R_{\mu\gamma}{}^{\eta\lambda}(R_{\nu\delta\eta\lambda}-2R_{\nu\eta\delta\lambda})+2R_{\mu\eta\gamma\lambda}(-R_{\nu\delta}{}^{\eta\lambda}+R_{\nu}{}^{\lambda}{}_{\delta}{}^{\eta})\right]+R_{\mu\nu}(\nabla_{\delta}\nabla_{\gamma}h^{\gamma\delta}-\Box h)\nonumber \\
 & +4R_{(\mu|}{}^{\gamma}\left[\nabla_{\gamma}\nabla_{|\nu)}h-\nabla_{\delta}\nabla_{\gamma}h_{|\nu)}{}^{\delta}+\Box h_{|\nu)\gamma}-\nabla_{\delta}\nabla_{|\nu)}h_{\gamma}{}^{\delta}\right]+2R_{\mu\gamma\nu\delta}\Box h^{\gamma\delta}-2R_{\nu\eta\gamma\delta}\nabla^{\eta}\nabla^{\delta}h_{\mu}{}^{\gamma}\nonumber \\
 & -(R_{\mu\delta\nu\eta}+R_{\mu\eta\nu\delta})(2\nabla^{\eta}\nabla_{\gamma}h^{\gamma\delta}-\nabla^{\eta}\nabla^{\delta}h)-2R_{\mu\eta\gamma\delta}\nabla^{\eta}\nabla^{\delta}h_{\nu}{}^{\gamma}+2R_{\nu\gamma\delta\eta}\nabla^{\eta}\nabla_{\mu}h^{\gamma\delta}\nonumber \\
 & +2R_{\mu\gamma\delta\eta}\nabla^{\eta}\nabla_{\nu}h^{\gamma\delta}+g_{\mu\nu}\left[h^{\gamma\delta}(-4R_{\gamma}{}^{\eta}R_{\delta\eta}+R_{\gamma\delta}R+R_{\gamma}{}^{\eta\lambda\xi}R_{\delta\eta\lambda\xi})-R\nabla_{\delta}\nabla_{\gamma}h^{\gamma\delta}\right.\nonumber \\
 & \left.+R\Box h-2R^{\gamma\delta}\left(\nabla_{\delta}\nabla_{\gamma}h-2\nabla_{\eta}\nabla_{\delta}h_{\gamma}{}^{\eta}+\Box h_{\gamma\delta}\right)+2R_{\gamma\eta\delta\lambda}\nabla^{\lambda}\nabla^{\eta}h^{\gamma\delta}\right]\nonumber \\
 & +R^{\gamma\delta}\left(2\nabla_{\delta}\nabla_{\gamma}h_{\mu\nu}-2\nabla_{\delta}\nabla_{\mu}h_{\nu\gamma}-2\nabla_{\delta}\nabla_{\nu}h_{\mu\gamma}+\nabla_{\mu}\nabla_{\nu}h_{\gamma\delta}+\nabla_{\nu}\nabla_{\mu}h_{\gamma\delta}\right)\nonumber \\
 & -\frac{1}{2}R\left(2\Box h_{\mu\nu}-2\nabla_{\gamma}\nabla_{\mu}h_{\nu}{}^{\gamma}-2\nabla_{\gamma}\nabla_{\nu}h_{\mu}{}^{\gamma}+\nabla_{\mu}\nabla_{\nu}h+\nabla_{\nu}\nabla_{\mu}h\right)\,.
\end{align}
As usual, suitable gauge fixing terms must be added to the quantum action, which remove the last terms in \eqref{Lich} and \eqref{Maxwellaa}, for diffeomorphism and gauge invariance, respectively.
\\
\\
With the second-order expansion of the action established, the quantum corrections to the partition function are encoded in the Euclidean functional integral. In the semi-classical approximation, this is expressed as:
\begin{equation}\label{functionalintegral}
Z=e^{-I[\bar\Phi]}\int D\phi e^{-\int\dd^5x\sqrt{\bar g}\phi^* \mathbb{O}\phi}=e^{-I[\bar\Phi]}Z_\text{1-loop}\ ,
\end{equation}
where the total field $\Phi=\bar\Phi+\phi$ is decomposed into a classical saddle-point configuration $\bar\Phi$ and a small quantum fluctuation $\phi$. For the sake of brevity, we suppress the explicit tensor structure of the fluctuations and assume the differential operator $\mathbb{O}$ to be self-adjoint with respect to the inner product $\int\dd^5x\sqrt{\bar g}\phi^* \mathbb{O}\phi=:(\phi,\mathbb{O}\phi)$.\\
The operator $\mathbb{O}$ possesses a complete set of normalizable, orthonormal eigenfunctions $\{u_i\}$, satisfying $(\phi,\mathbb{O}\phi)=\sum_i\lambda_ic_i^2$. By expanding the fluctuations $\phi(x)=\sum_i c_i u_i(x)$, the quadratic form in the exponent diagonalizes as $(\phi,\mathbb{O}\phi)=\sum_i\lambda_ic_i^2$. Consequently, the functional integral
\eqref{functionalintegral}, reduces to a product of Gaussian integrals over the expansion coefficients $c_i$, yielding 
\begin{equation}\label{Zgen}
Z\sim e^{-I[\bar\Phi]}\frac{1}{\sqrt{\det\mathbb{O}}}=e^{-I[\bar\Phi]}\prod_i\frac{1}{\sqrt{\lambda_i}}\ .
\end{equation}
While we have focused on bosonic fields, gauge invariance necessitates the inclusion of Faddeev-Popov ghosts associated with diffeomorphism and $U(1)$ gauge symmetries. These formal considerations are subtle due to the existence of zero-modes for the operator $\mathbb{O}$, which leads to divergences that must be regularized. Specifically, the zero-modes of $\mathbb{O}$ on the extremal black hole geometry are those responsible for the logarithmic contribution to the entropy, as we will see below.\\
This mechanism can be understood schematically as follows. The near-extremal, near-horizon geometry \eqref{NHNEXT} is decomposed into the AdS$_2\times S^3$ background, plus a perturbative part which is linear in the temperature $T$. Therefore, the operator $\mathbb{O}$ inherits a similar splitting, namely $\mathbb{O}=\mathbb{O}_{{\rm AdS_2}\times S^3}+\delta\mathbb{O}$, where $\delta\mathbb{O}$ is also linear in $T$. The structure is further reflected in the eigenvalues $\lambda_i=\lambda_i^{{\rm AdS_2}\times S^3}+\delta\lambda_i$, where at first order:
\begin{equation}\label{definiciondeltalambda}
\delta\lambda_i=(u_i,\delta\mathbb{O}u_i)\ .
\end{equation}
Here, the eigenfunctions $u_i$ correspond to those of the unperturbed operator, as usual in first-order perturbation theory. It follows from \eqref{Zgen} that the one-loop partition function acquires a logarithm term in the temperature exclusively from the zero modes of the AdS$_2\times S^3$ background, since in such case
\begin{equation}
\log Z_\text{1-loop}\sim -\frac{1}{2}\sum_i\log(\lambda_i^{{\rm AdS_2}\times S^3}+\delta\lambda_i)\sim-\frac{1}{2}\sum_i^{(0)}\log(\delta\lambda_i)\ ,
\end{equation}
where the sum $\sum_i^{(0)}$ runs over the zero modes of $\mathbb{O}$ on the extremal near-horizon geometry AdS$_2\times S^3$ (see \cite{Iliesiu:2022onk,Karan:2022dfy,Banerjee:2023quv, Kapec:2023ruw,Rakic:2023vhv,Maulik:2024dwq,Maulik:2025phe, Blacker:2025zca,Acito:2025hka, PandoZayas:2026vbg}). In summary, the computation of the logarithmic correction to the entropy of near-extremal black holes in the Einstein-Gauss-Bonnet theory, also requires to compute the correction to the eigenvalue of the zero-modes induced by finite temperature effects, namely to compute $\delta\lambda_i$ defined in \eqref{definiciondeltalambda}. In what follows we proceed to perform this computation. Notice that now both the operator $\mathbb{O}_{{\rm AdS_{2}}\times S^3}$ and $\delta\mathbb{O}$ depend non-trivially on the Gauss-Bonnet coupling $\alpha$ (see the formulae of the previous section).\\
In particular, the operator in \eqref{functionalintegral} is given by
\begin{align}
    -\int \dd^5 x \sqrt{\overline{g}} \phi \mathbb{O} \phi = - \int \dd^5 x \sqrt{\overline{g}} \left[ h_{\mu \nu} \mathsf{D}^{\mu \nu, \rho \sigma} h_{\rho \sigma} + 
    h_{\mu \nu} \mathsf{D}^{\mu \nu, \rho} a_{\rho} +
    a_{\mu} \mathsf{D}^{\mu, \rho \sigma} h_{\rho \sigma} + 
    a_{\mu} \mathsf{D}^{\mu, \rho} a_{\rho}
    \right] \, .
\end{align}
However, we have proven that $h_{\mu \nu} \mathsf{D}^{\mu \nu, \rho} a_{\rho} = a_{\mu} \mathsf{D}^{\mu, \rho \sigma} h_{\rho \sigma}$ for appropriate boundary conditions. The differential operator for the bilinear in the metric perturbation is 
\begin{align} \label{hDhoperator}
    h_{\mu \nu} \mathsf{D}^{\mu \nu, \rho \sigma} h_{\rho \sigma} &= - \frac{1}{\kappa^{2}}
    h_{\rho \sigma} \left[\frac{1}{4}g^{\mu \rho}g^{\nu \sigma}\Box-\frac{1}{8}g^{\rho \sigma}g^{\mu\nu}\Box-\frac{1}{2}R^{\mu \rho \sigma\nu}+\frac{1}{2}g^{\sigma\mu}R^{\rho \nu}\right.\nonumber\\
 & \quad-\frac{1}{2}g^{\mu\nu}R^{\rho \sigma}-\frac{1}{4}g^{\mu \rho}g^{\nu\sigma}R+\frac{1}{8}Rg^{\rho \sigma}g^{\mu\nu}  -\frac{\Lambda}{4}g^{\rho\sigma}g^{\mu\nu}+\frac{\Lambda}{2}g^{\mu\rho}g^{\sigma\nu} \nonumber\\
 & \quad + \frac{1}{8}F_{\lambda\delta}F^{\lambda\delta}\left(2g^{\rho\mu}g^{\sigma\nu}-g^{\rho\sigma}g^{\mu\nu}\right)-F^{\mu\rho}F^{\nu\sigma}-2F^{\rho\lambda}F^{\mu}{}_{\lambda}g^{\sigma\nu}+g^{\mu\nu}F^{\rho}{}_{\lambda}F^{\sigma\lambda} \nonumber\\
 & \left. \quad-\frac{\alpha}{4}g^{\rho\sigma}H^{\mu\nu}+\alpha g^{\mu\rho}H^{\sigma\nu}-\frac{\alpha}{2}\mathsf{P}[H]^{\mu\nu,\rho\sigma} \right]h_{\mu\nu} \, .
\end{align}
The interaction term operator and bilinear in the $U(1)$ field are:
\begin{align}
    h_{\mu\nu}\mathsf{D}^{\mu\nu,\rho}a_{\rho} & = \frac{2}{\kappa^{2}}h_{\mu\nu}\left(g^{\mu\nu}F^{\alpha\beta}+4g^{\mu\lambda}g^{\nu[\alpha}F_{\lambda}{}^{\beta]}\right)\nabla_{\alpha}a_{\beta}\,, \label{interaction_operator} \\
    a_{\mu} \mathsf{D}^{\mu, \rho} a_{\rho} &= -\frac{2}{\kappa^{2}}a_{\mu}(g^{\mu\lambda}\Box-R^{\mu\lambda})a_{\lambda}\,. \label{U1 operator}
\end{align}

The computation of zero-modes coming from the gravitational sector can be performed as follows. As in General Relativity, Einstein-Gauss-Bonnet theory is invariant under diffeomorphisms, a symmetry that at perturbative level acts on the fluctuation $h_{\mu\nu}$ of the metric above a background $\bar{g}_{\mu\nu}$ via the Lie derivative as $\delta_\xi h_{\mu\nu}=\mathcal{L}_\xi \bar{g}_{\mu\nu}$, along a vector field $\xi\in T_p(\mathcal{M})$. Notice that no ambiguity on the number of degrees of freedom propagated by the Einstein-Gauss-Bonnet gravitons is present, since we are eventually interested in dealing with the Gauss-Bonnet coupling $\alpha$, perturbatively \cite{Zwiebach:1985uq}. The gauge fixing condition allows to construct zero-modes of the operator $\mathbb{O}$ on the extremal, near-horizon geometry, by considering normalizable $(h_{\mu\nu},h_{\mu\nu})=1$, transverse $\bar{g}^{\mu\nu}\bar{\nabla}_\mu h_{\nu\alpha}=0$ and traceless $\bar{g}^{\mu\nu}h_{\mu\nu}=0$ fluctuations of the form $h_{\mu\nu}=\mathcal{L}_\xi \bar{g}_{\mu\nu}$. These normalizable modes are eventually constructed from non-normalizable vector fields $\xi$, and a diffeomorphism along them acts non-trivially on the asymptotic region of the  near-horizon extremal geometry (see e.g. \cite{Rakic:2023vhv}). It can be shown that there are two types of zero-modes of the operator $\mathbb{O}_{{\rm AdS_{2}}\times S^3}$, which contribute to the logarithmic terms we are looking for. In addition to these gravitational modes, there are modes associated to the $U(1)$ gauge field that do contribute to the logarithmic terms that we also consider. In what follows we deal of each family of modes separately.
\\
\paragraph{\bf Tensor modes:} These modes are constructed from solutions of the scalar equation on ${\rm AdS_2}$ 
\begin{equation}
( \bar{\square}_{\rm AdS_2}+R_{\rm AdS_2})B(\tau,\eta)=0\ , \label{box-R-F-equation}
\end{equation}
leading to the tensor modes
\begin{equation}
h_{\mu\nu}=\mathcal{L}_\xi \bar{g}_{\mu\nu}\quad\text{ with }\quad \xi = \xi^{\flat}_\mu \bar{g}^{\mu \nu} \partial_\nu \,,\quad  \xi^{\flat}=\star_{\rm AdS_2} \dd B\ . \label{tensor_pert}
\end{equation}
The Hodge star is computed with the AdS$_2$ metric \eqref{metric-ads} and $R_{\rm AdS_2} $ stands for its curvature scalars. This is so because the condition $h_{\mu \nu} \bar{g}^{\mu \nu} = 0$ implies that $\dd \star_{\rm AdS_2} \xi^\flat = 0$. By using Poincar\'e's lemma, it follows that the most general solution is the last equation in \eqref{tensor_pert} for some scalar function $B$. Then, the transverse condition implies \eqref{box-R-F-equation}.\\
For the present case, the normalized perturbation is of the form
\begin{align}
h_{\mu\nu}^{(n)}\dd x^{\mu}\dd x^{\nu} & =\frac{\ri e^{\ri n\tau}\ell_{\mathrm{AdS}}\sqrt{|n|(n^{2}-1)}}{2\pi^{3/2}r_{0}^{3/2}}\tanh^{|n|}\left(\frac{\eta}{2}\right)\left(\ri\dd\tau+\frac{\dd\eta}{\sinh\eta}\right)^{2}\,. \label{htensor}
\end{align}
These tensor zero-modes of the extremal near-horizon geometry are lifted at finite, small temperature, leading to
\begin{align}\label{deltatensor}
    \delta \lambda_{n}^{\rm tensor}&=\left(h_{\mu\nu}^{(n)},\delta\mathbb{O}h_{\mu\nu}^{(n)}\right)= \int \dd^5x \sqrt{\bar{g}}  h^{(n)*}_{\mu \nu} \delta \mathsf{D}^{\mu \nu , \rho \sigma} h_{\rho \sigma}^{(n)}= \frac{|n| T }{T_{\rm tensor}} + \mathcal{O}(T^2) \, , \\
    T_{\rm tensor} &= \frac{64 G r_0^3}{3 \pi  \left(4 \alpha +r_0^2\right)} \label{Ttensor} \,,
\end{align}
with $|n|=2,3,\ldots$, since we require the mode to be normalizable on the near-horizon, extremal geometry. As in GR, the correction to the zero-modes is linear in the mode number $n$. This leads to the well known $3/2$ factor in front of the logarithmic correction to the extremal black hole entropy at small temperature and fixed charge. Notice that for positive $\alpha$, as predicted by String Theory, the coefficient $T_{\rm tensor}$ in \eqref{Ttensor} is manifestly positive and corresponds to a characteristic scale of the tensor modes. Also, for fixed $r_0$, the correction is linear in the GB coupling $\alpha$.
\\
\paragraph{\bf Vector modes:} These modes are constructed from solutions of
\begin{equation}
\bar{\square}_{\rm AdS_2} V(\tau,\eta)=0 \, , \label{box ads V}
\end{equation}
as the building block for the metric fluctuation
\begin{align}
    h^{(n,m)}_{\mu \nu} \dd x^\mu \dd x^\nu = \dd V_n \otimes \zeta_m + \zeta_m \otimes \dd V_n = 2  \zeta_{(m} \dd V_{n)}  \,,
\end{align}
 where $n$ labels the $n$-th mode of the box operator in \eqref{box ads V}, and $\zeta_{m}$ represents a given $m$-th one-form, whose contravariant components lead to a Killing vector of the 3-sphere for each $m=1,\dots, 6$. For the metric \eqref{metric-sphere}, the isometry algebra is $\mathfrak{so(4)}$, and we can consider a complex basis for $\mathfrak{so(4)}\cong\mathfrak{su(2)\times su(2)}$:
\begin{align}
\zeta_{1} & =e^{\ri\phi}(\dd\theta-\ri\sin\theta\dd\psi)\,, &  & \zeta_{4}=e^{\ri\psi}(\dd\theta-\ri\sin\theta\dd\phi)\,,\\
\zeta_{2} & =e^{-\ri\phi}(\dd\theta+\ri\sin\theta\dd\psi)\,, &  & \zeta_{5}=e^{-\ri\psi}(\dd\theta+\ri\sin\theta\dd\phi)\,,\\
\zeta_{3} & = \ri \sqrt{2}(\dd\phi+\cos\theta\dd\psi)\,, &  & \zeta_{6}= \ri \sqrt{2}(\dd\psi+\cos\theta\dd\phi)\,,
\end{align}
where the generators of the Cartan subalgebra are $\zeta_{3},\zeta_{6}$. They satisfy the Maurer-Cartan equations $\sqrt{2} \dd\zeta_{1}+\zeta_{1}\wedge\zeta_{3}=0,$ $\sqrt{2}\dd\zeta_{2}-\zeta_{2}\wedge\zeta_{3}=0,$ $\sqrt{2} \dd\zeta_{3}+\zeta_{1}\wedge\zeta_{2}=0$, mutatis mutandis for the second $\mathfrak{su(2)}$ copy.

Regularity of the function $V(\tau,\eta)$ at the origin implies that the most general solution is $V_n(\tau,\eta)=A_n e^{in\tau }\tanh^{|n|}\left(\frac{\eta}{2}\right)$ with $A_n$ are normalization constants. The vector perturbation in this normalization is
\begin{align}
    h^{(m,n)}=  \frac{1}{8\sqrt{\pi^3 |n| r_0}} e^{in\tau }\tanh^{|n|}\left(\frac{\eta}{2}\right) \left(\ri n\dd\tau+\frac{|n|}{\cosh\eta}\dd\eta\right)\otimes_{s}\zeta_m.\label{hvector}
\end{align}
The operation on the right-hand side is the symmetric tensor product, defined without the conventional factor of 1/2. For these gravitational vector modes, the corrections are independent of the Killing vector labeled by $m$ and are given by
\begin{align}
    \delta \lambda_{m,n}^{\rm vector}&=\left(h^{(m,n)},\delta\mathbb{O}h^{(m,n)}\right)= \int \dd^5x \sqrt{\bar{g}}  h^{(m,n)*}_{\mu \nu} \delta \mathsf{D}^{\mu \nu , \rho \sigma} h_{\rho \sigma}^{(m,n)} \,, \\
    &=
    \frac{ |n| T}{T_{\rm vector}} + \mathcal{O}(T^2)\,,\\
    T_{\rm vector} &=  \frac{48 G r_0^3 \left(4 \alpha +r_0^2\right) \left(\Lambda  r_0^2-2\right)}{\pi\left(-48 \alpha ^2+24 \alpha ^2 \Lambda  r_0^2+4 \alpha  \Lambda  r_0^4-12 \alpha  r_0^2+\Lambda  r_0^6-3 r_0^4\right)} \,.
\end{align}
In this case, $|n|=1,2,\ldots$. For $\alpha>-r_0^2/4$ the correction is always positive for any $\Lambda<0$, while for $\alpha<-r_0^2/4$ is always negative. The expansion for small Gauss-Bonnet coupling is as follows
\begin{align}
    \delta \lambda_{m,n}^{\rm vector} = \frac{|n| \pi T}{G r_0^3} \left[ \frac{r_0^2 (3 - r_0^2 \Lambda)}{2-r_0^2 \Lambda} + \frac{\alpha^2}{2 r_0^2} - \frac{2 \alpha^3}{r_0^4} + \mathcal{O}(\alpha^4/r_0^6 ) \right] \,.
\end{align}
Interestingly, the linear correction in $\alpha$ is absent from the final expression.
\\
\paragraph{\bf $U(1)$ modes:} The $U(1)$ gauge modes $a_\mu(x)$ are zero-modes of the operator \eqref{U1 operator} and must satisfy the gauge fixing condition $\bar{\nabla}_\mu a^\mu=0$. Pure gauge configurations (i.e. $a_\mu = \partial_\mu V$) are automatically zero-modes of the operator \eqref{U1 operator}, then the gauge fixing conditions implies $\bar{\Box} V =0$. Given the direct product of the space, the relevant modes are non-trivial along the ${\rm AdS_2}$, which coincide with the solutions of \eqref{box ads V}. Hence, these modes are labeled by the integer numbers $n$ and are given by
\begin{equation}
    a^{(n)}_\mu \dd x^\mu = \frac{1}{2\sqrt{|n| \pi^3 r_0^3}} e^{\ri n\tau}\tanh^{|n|}\left(\frac{\eta}{2} \right) \left(\ri n \, \dd\tau+\frac{|n|}{\sinh\eta}\dd\eta\right)  \,.\label{agauge}
\end{equation}
Normalizability of the modes require $|n|=1,2,\ldots$. They are orthogonal with respect to the previously defined inner product $(a^{(n)},a^{(m)})=\delta_{n,m}$. Here, the small temperature correction to the $U(1)$ gauge zero-modes is given by 
\begin{align}
   \delta \lambda_{n}^{\ U(1)} &= \left(a^{(n)}_\mu,\delta\mathbb{O}a^{(n)}_\mu\right)= \int \dd^5x \sqrt{\bar{g}} a_{\mu}^* \delta \mathsf{D}^{\mu , \rho} a_{\rho} = 
   \frac{|n| T}{T_{U(1)}}+\mathcal{O}(T^2)\,,\\
   T_{U(1)} &=  -\frac{12 G r_0 \left(4 \alpha +r_0^2\right) \left(\Lambda  r_0^2-2\right)}{\pi  \left(-24 \alpha +16 \alpha  \Lambda  r_0^2+\Lambda  r_0^4\right)}\, .
\end{align}
This perturbative correction only exists when $\alpha$ or $\Lambda$ are non-vanishing. The correction is also always negative for $\Lambda < 0$ and $\alpha>0$. Notice that the contribution of these modes is negative in General Relativity with a negative cosmological constant as well, and vanish when the bulk cosmological constant vanishes. This can be directly seen from equation (4.31) of \cite{Blacker:2025zca}. Extrapolating our result, if we allow $\Lambda>0$, there exists a region in which the correction is positive (see \cite{Blacker:2025zca}). In our case, the expansion of the correction for $\alpha/r_0^2\ll0$ is 
\begin{equation}
    \delta \lambda_{n}^{ U(1)}= \frac{\pi |n| T}{G r_0}\left[-\frac{\Lambda  r_0^2}{12 \left(\Lambda  r_0^2-2\right)} -\frac{\alpha }{r_0^2} +\frac{4 \alpha ^2}{r_0^4} -\frac{16 \alpha ^3}{r_0^6} + \mathcal{O}(\alpha^4/r_0^6 )\right] \, .
\end{equation}
It is worth pointing out that the $U(1)$ gauge zero-modes are also annihilated by the interaction operator \eqref{interaction_operator}.Thus, to obtain a non-vanishing interaction contribution, one should consider perturbations that are not generated by a gauge transformation.
\\
\paragraph{\bf Possible existence of other modes:} 
In 3-dimensions there exist non-trivial tensor modes that are eigenvectors of the Laplace-Beltrami operator. Since we are working in a background whose near-horizon region contains a 3-sphere, it is worth looking for the tensorial perturbations on the 3-sphere that could contribute to the path integral. Let us show that, if in addition we require that the perturbations are transverse-traceless and are generated by a gauge transformation, then they do not exist for any $p$-dimension sphere. Considering a maximally symmetric $p$-dimensional manifold $N$ with metric $\gamma_{ij}$, curvature $K$, covariant derivative $D_i$ and curved indices $i,j,\dots$. If we consider that the perturbations are generated by a gauge transformation $h_{ij} = 2 D_{(i}\xi_{j)}$ and we impose that they are traceless $h_{ij}\gamma^{i j}=0$, and transverse $D_{i}\gamma^{i j}=0$, then the resulting equation is $D_i D^i \xi^j = -K(p-1) \xi^j$. Hence, $\xi$ is a vector harmonic on $N$. If we restrict $N=S^p$, then \cite{Higuchi:1986wu,Kodama:2003jz,Ishibashi:2011ws} the eigenvalues of the vector harmonics\footnote{For the general construction see Theorem 3.2 of \cite{Ishibashi:2011ws}.} are fixed leading to the following relation 
\begin{equation}
    p = l(l+p-1) \,
\end{equation}
where $l=1,2,\dots$. The only solution to this equation is found to be $l=1$, which leads to $p(p+1)/2$ different harmonics. These vector harmonics coincide with Killing vectors of the sphere leading to a vanishing perturbation $h_{ij}$.\\
\\
Taking all these contributions into account, and in order to obtain the total contribution to the one-loop partition function of the tensor modes \eqref{htensor}, gravitational vector modes \eqref{hvector} and $U(1)$ gauge modes \eqref{agauge}, we need to use the regularized products
\begin{equation}\label{regularizadas}
\prod\limits_{n=2}^{\infty}\frac{\xi}{n}=\frac{1}{\xi^{3/2}\sqrt{2\pi}}\ \ \text{and} \ \ \prod\limits_{n=1}^{\infty}\frac{\xi}{n}=\frac{1}{\sqrt{2\pi\xi}}\ ,
\end{equation}
respectively. Appropriately accounting for the factor of $2$ emerging from the positivity and negativity of the mode numbers in each case, one shows that the contribution of the zero-modes on the near-horizon near-extremal configuration, to the 1-loop partition function of the near-extremal black holes are given by
\begin{equation}
\log{Z^{(0)}_\text{1-loop}}=\frac{3}{2}\log{\frac{T}{T_{\rm tensor}}}+\frac{6}{2}\log{\frac{T}{T_{\rm vector}}}+\frac{1}{2}\log{\frac{T}{T_{U(1)}}}+\ldots\ , \label{famous-logT}
\end{equation}
where the ellipsis stands for numerical values. Notice that we have retained the physical constant within the argument of the logarithms. The factor of $6$ in the second term, corresponding to the gravitational vector modes, arises because each of the six Killing vectors of $S^3$, generates a tower of modes, each contributing a factor of $1/2$ to the total coefficient.
\section{Conclusions}

In this work, we have computed the one-loop quantum corrections to the thermodynamics of near-extremal, asymptotically AdS charged black holes in five-dimensional Einstein-Gauss-Bonnet gravity. Our analysis is based on the structure of the generalized Lichnerowicz operator governing quadratic fluctuations around the near-horizon geometry, and on the identification and lifting of its zero-modes under a small temperature deformation.

Working in the canonical ensemble at fixed charge, we exploited the decoupling of the near-horizon region in the low-temperature limit, where the geometry reduces to an $\mathrm{AdS}_2 \times S^3$ background with $\alpha$-dependent radii. In this regime, the relevant contribution to the one-loop partition function arises entirely from zero-modes of the extremal background, whose eigenvalues are lifted linearly in the temperature. This mechanism leads, as in General Relativity, to logarithmic corrections to the entropy.

Our main result is that the one-loop partition function exhibits a logarithmic dependence on the temperature of the form
\begin{equation}
\log Z^{(0)}_{\text{1-loop}} \sim \frac{3}{2}\log\frac{T}{T_{\text{tensor}}}
+ \frac{6}{2}\log\frac{T}{T_{\text{vector}}}
+ \frac{1}{2}\log\frac{T}{T_{U(1)}} + \cdots,
\end{equation}
leading to a total contribution $5 \cdot \log T$ at low temperatures. This result matches general expectations from the structure of zero-modes and reflects the contribution of gravitational tensor modes, vector modes associated with the isometries of $S^3$, and $U(1)$ gauge modes.

A key outcome of our analysis is that, while the overall coefficient of the logarithmic correction remains unchanged with respect to General Relativity, the characteristic scales $T_{\text{tensor}}$, $T_{\text{vector}}$, and $T_{U(1)}$ acquire a nontrivial dependence on the Gauss--Bonnet coupling $\alpha$. In particular, tensor modes receive corrections linear in $\alpha$, vector modes exhibit a more subtle dependence with no linear contribution in the small-$\alpha$ expansion, and the $U(1)$ sector depends sensitively on both $\alpha$ and the bared cosmological constant. These features reflect the deformation of the fluctuation operator induced by higher-curvature terms.

Our results also reinforce the persistence of the near-extremal entropy puzzle: the coexistence of a large extremal entropy with a finite energy gap $E_{\text{gap}}$, which in this case depends explicitly on $\alpha$. The emergence of logarithmic corrections is essential for consistency with the thermodynamic expansion, yet the microscopic origin of the large ground-state degeneracy remains unclear in Einstein-Gauss-Bonnet gravity. Notice that the solution we have considered can be embedded in a higher curvature deformation of minimal gauged $\mathcal{N}=2$ supergravity by the inclusion of the Gauss-Bonnet term, since the electric ansatz for the gauge field makes the contribution of the Chern-Simons term, $A\wedge F\wedge F$, to vanish identically. 

It would be interesting to extend this analysis to rotating solutions, as well as to explore a possible holographic interpretation of the $\alpha$-dependence of the characteristic scales. Regarding the former, as stated in the introduction, in spite of the existence of scattered rotating solutions \cite{Anabalon:2009kq,Cvetic:2016sow,Dehghani:2002wn,Dehghani:2003ea,Dehghani:2006dh,Dehghani:2006cu,Kim:2007iw,Brihaye:2008kh,Brihaye:2013vsa,Konoplya:2020fbx,Anabalon:2024abz}, none of them describe an exact, analytic, rotating solution with two angular momenta, for arbitrary values of the coupling $\alpha$, and that is the reason why in this work we have focused on the static, near-extremal charged case.

\acknowledgments
We thank Nicolás Grandi, Gastón Giribet and Alejandra Castro for comments. The work of A.A and M.C. is partially funded by DGAPA-UNAM grant IG101326. This work is supported in part by the FONDECYT grants 1230853, 1242043, 1250133, 1262452 and  126241. The work of AA is supported in part by the FAPESP grant 2024/16864.

\hypersetup{linkcolor=blue}
\phantomsection 
\addtocontents{toc}{\protect\addvspace{4.5pt}}
\providecommand{\href}[2]{#2}
\end{document}